\documentstyle[aps,prl,eqsecnum,preprint,tighten]{revtex}
\begin{document}
\title{The edge theory of ferromagnetic quantum Hall states}
\author{M.V.\ Milovanovi\'{c}} 
\address{Physics Department, Technion, Haifa 32000, Israel}
\date{\today}
\maketitle
\begin{abstract}
We propose an effective low-energy theory for ferromagnetic Hall states. It
describes the charge degrees of freedom, on the edge, by a (1 + 1)
dimensional chiral boson theory, and the spin degrees of freedom by the (2 + 1)
dimensional quantum ferromagnet theory in the spin-wave approximation. The
usual chiral boson theory for spinless electrons is modified to include the
charge degrees of freedom with spin. Our total, bulk plus edge, effective
action is gauge invariant and we find a generalized "chiral anomaly" in
this case. We describe two, charged and neutral, sets of
edge spin-wave solutions. The spreading of these waves is much larger than the
one for the charge (edge) waves and they have  linear dispersion
relations. 
\end{abstract}


\section{Introduction}
\label{introduction}
The bulk properties of the quantum Hall systems at filling fraction
$1/m$, $m =$ odd, in the presence of low magnetic fields have been
subject of many theoretical and experimental investigations in recent
years. The spin degree of freedom plays an important role in these
systems. Here we focus on properties of the boundary of these systems,
which, in a special way, reflect bulk properties.
In the spinless case this reflection was already described by Wen
\cite{wen}.
The low-energy (bulk) physics of these systems is identical to the one
of 2D quantum ferromagnets with spin waves as excitations. Due to
exchange the spins of electrons in the ground state are all aligned in
the same direction and the lowest lying excitations are one-spin-flip
(spin wave) excitations which leave the charge of the system unchanged.
The lowest lying charged excitations are topologically non-trivial
skyrmion excitations \cite{sond} for which a local change in the charge density that
characterizes them, is accompanied by a local change in the spin
density. This scenario, in which a finite number of overturned spins
follows the creation of the charged excitations is supported by
experimental findings \cite{barr}.

On the other hand the physics of the boundary of quantum Hall systems
without the spin degree of freedom is well understood \cite{wen}. 
In fact for any
quantum Hall system, including the one which edge physics we would like
to understand, it is expected that the charge dynamics is reduced to the
edge of the system. This is maybe an oversimplification with respect to a
general experimental situation where sharp edges (i.e. edges with steep
confining potential) are not always present. Still the effective
one-space-dimensional edge theory - chiral boson theory of quantum Hall
systems has received considerable experimental support in recent years.
 
Here we present a low-energy effective theory of quantum Hall
ferromagnetic systems which describes the charge degrees of freedom
restricted to live only on the edge and lowest lying excitations of the
bulk - neutral spin waves. We show that, under special conditions (and as
solutions of the theory), edge spin waves exist, whose characteristic
width is smaller than the one of bulk spin waves, which spread
throughout the whole system.  These excitations, edge spin waves, are
characterized by gaps which are smaller than the Zeeman gap and linear
dispersion relations. We find two classes of these waves, wich we call,
charged and neutral edge spin waves. One way to induce the charged edge
waves is to subtract or add some charge to the edge. By a redistribution
of the charge and, simultaneously, spin of the system on the edge (in
the manner of spin textures as described first in \cite{sond}) 
neutral edge spin
waves are possible. 
\section{Effective low-energy field theory with charge degrees of
freedom on the edge and spin waves}
In this section, we will first rederive the edge theory for spinless
electrons \cite{wen} using the dual form of the Chern-Simons field-theory
description of quantum Hall systems (at filling fractions $1/m$ where
$m$ is an odd integer . Then we
will use the dual form of the Chern-Simons formulation of the
ferromagnetic quantum Hall systems (at the same filling fractions with
the spin degree of freedom taken into account) to derive a low energy
effective theory which describes not only the edge of these systems, but
also lowest lying excitations in their bulk - spin waves.
At the end we will demonstrate the gauge invariance of our total, bulk
plus
edge, action when perturbing external electromagnetic fields are present
in it.
\subsection{The edge theory of spinless quantum Hall systems}
The systems that we consider are at filling fractions $1/m$ where $m$ is
an odd integer. We start with the dual formulation of the Chern-Simons
effective description of these systems. In the Chern-Simons formulation
the problem of the 2D electron system in magnetic field is mapped to a
problem of bose liquid with a long-range interaction described by a
ststistical
gauge  field. The vortex excitations of the bose fluid
correspond to the quasiparticle excitations of quantum Hall systems. In
the dual form the vortex excitations (i.e. fluxes of the gauge field in
the preceding formulation) are viewed as particles, and the
charge
-current density becomes flux of some new gauge field. The first
terms of the Lagrangian density in the dual form are:
\begin{equation}
{\cal L}
 = -\frac{m}{4 \pi} \epsilon^{\mu \nu \lambda} A_{\mu}
\partial_{\nu} A_{\lambda} + \frac{1}{2 \pi} \epsilon^{\mu \nu
\lambda} A^{ext}_{\mu} \partial_{\nu} A_{\lambda} -  \label{lagr}
{\cal J}^{v}_{\mu}
 A^{\mu} -  \frac{\lambda}{4} F_{\mu \nu} F^{\mu \nu}
\end{equation}
The vector potential $A_{\mu},\; \mu = 0,1,2$ represents the newly
introduced gauge field, which enters the definition of the charge
current-density $J_{\mu},\; \mu = 0,1,2$:
\begin{equation}
J_{\mu} = \epsilon_{\mu \nu \lambda} \frac{1}{2 \pi} \partial^{\nu}
A^{\lambda}. \label{basic}
\end{equation}
$F_{\mu \nu} = \partial^{\mu} A^{\nu} - \partial^{\nu} A^{\mu}.$
The vector potential $A_{\mu}^{ext}, \mu = 0,1,2$ describes the external
electromagnetic field, and two first terms  when $A_{\mu}$ is
``integrated out'' give the basic Hall response of the system i.e. the
Hall conductance equal to $ \frac{1}{m} \frac{e^{2}}{h}$. The third term
couples the vortex current-density $ {\cal J}_{\mu}^{v},\; \mu = 0,1,2$,
\begin{equation}
{\cal J}_{\mu}^{v}
 = \frac{1}{2 \pi i} \epsilon_{\mu \nu \lambda} 
\partial^{\nu} \partial^{\lambda} \alpha     \label{vorcur}
\end{equation}
to the gauge field $A_{\mu}$ (according to the interchanged roles of fluxes
and particles). In \ref{vorcur} $\alpha$ is the phase of the bose field
in the former (non-dual) formulation and the vortex current-density is
non-zero only if $ \alpha$ is a nonanalytic function of coordinates i.e.
$ \partial^{\nu} \partial^{\lambda} \alpha \neq \partial^{\lambda}
\partial^{\nu} \alpha $ for some $\lambda$ and $\nu$.

If our system has a boundary the action for a general gauge
transformation $A_{\mu} \rightarrow A_{\mu} + \partial_{\mu} \Lambda$
for which $\Lambda$ is not zero at the boundary is not gauge invariant
(i.e. charge conserving). Any electric field along the boundary will
produce a current normal to the boundary because of the nonzero value of
the Hall conductance.  But   with 
restricted gauge transformations for which $\Lambda=0$ on the boundary we
may use this action to
derive, as was already done in \cite{wen}, the kinetic term of the edge theory
Lagrangian. These gauge transformations 
 describe a well-defined
boundary problem, in which some of the physics of the quantum Hall
systems due to perturbing external electromagnetic fields is not present. Namely,  by making all previously gauge
dependent quantities on the edge gauge independent i.e. physical, we are
deriving
 an effective edge theory
that gives the right physics when charge exchange between the bulk and
edge is absent \cite{ston,hald}. (We defer description of the total action with
perturbing electromagnetic fields which is explicitly gauge invariant to
subsection II C.)

First we neglect the last term in \ref{lagr} (as a higher order term in
derivatives), 
regard the equation of motion for $A_{0}$ as a constraint, and take
$A_{0} = 0$.
 The constraint is
\begin{equation}
\frac{1}{2 \pi}(m \vec{\nabla} \times \vec{A} - \vec{\nabla} \times
\vec{A}^{ext}) = - {\cal J}_{0}^{v}
\end{equation}
simply saying that any deviation in the charge density of the system is
due to the creation of vortices. The solution of this equation, up to a
gauge transformation is
\begin{equation}
\delta A_{a} = A_{a} - \frac{1}{m} A_{a}^{ext} = - \frac{1}{m}
\partial_{a} \alpha.   \label{solu} 
\end{equation}
We assume that there are no vortices in the bulk of the system so that
$\alpha$ is analytic. When we plug in the solution \ref{solu} in the
remaining term of the Lagrangian (we do not consider the term with
$A_{\mu}^{ext}$):
\begin{equation}
\Delta {\cal L}_{\rm eff} = - \frac{m}{4 \pi} \epsilon_{a 0 b} A_{a}
\partial_{0} A_{b}
\end{equation}
we get, up to a total time derivative and with an assumption that
$A_{\mu}^{ext}$ is time independent, that
\begin{equation}
{\cal L}_{\rm eff}(x,y,t)
 = \frac{1}{4 \pi m} \epsilon_{a b} \partial_{b}
(\partial_{a} \alpha \partial_{0} \alpha).
\end{equation}
This total divergence can be translated into a surface term:
\begin{equation}
{\cal L}_{\rm eff}(x,t)
 = \frac{1}{4 \pi m} \partial_{x} \alpha \partial_{0}
\alpha,  \label{surf}
\end{equation}
exactly the kinetic term of the chiral boson theory, if we consider the
system to be defined in the lower half-plane with $y=0$ as a boundary.
We get only the kinetic term of the edge theory because we started from
 the theory in which we neglected  terms that bring dynamics. We
might expect that the following term in small momentum expansion on the
edge is
\begin{equation}
- \frac{v}{4 \pi m} \partial_{x} \alpha \partial_{x} \alpha
\end{equation}
with a nonuniversal coupling $v$. This term gives dynamics to the edge
theory and together with \ref{surf} makes the chiral boson theory Lagrangian
density. Thus, we can conclude that the field $\alpha$ - phase of the
bosonic field in the standard Chern-Simons formulation on the boundary
plays the role of the chiral boson field.

\subsection{Effective low-energy field theory of ferromagnetic quantum
Hall states}
The Lagrangian density of the Chern-Simons theory for ferromagnetic
quantum Hall states in the dual form is \cite{leeka,stone}:
\begin{eqnarray}
 \cal{L} = & & - \frac{m}{4 \pi} \epsilon^{\mu \nu \lambda} A_{\mu}
\partial_{\nu} A_{\lambda} + \frac{1}{2 \pi} \epsilon^{\mu \nu \lambda}
A_{\mu}^{ext} \partial_{\nu} A_{\lambda} - {\cal J}_{\mu} A^{\mu} \nonumber
\\ \label{lagrs}
  & & - \frac{\rho_{s}}{2} ( \vec{\nabla} \vec{n})^{2} +
\frac{\Delta}{2} n_{3} - \frac{\lambda}{4} F^{\mu \nu} F_{\mu \nu}.
\end{eqnarray}
Again the charge current-density as a function of the statistical gauge
field $A^{\mu}$ is given by the relation \ref{basic}. Now quasiparticle
current-density consists of two contributions: vortex and skyrmion:
${\cal J} = {\cal J}_{\mu}^{v} + {\cal J}_{\mu}^{s}$. As before vortex
excitations do not change the spin configuration of the ferromagnetic
Hall states. Skyrmions, on the other hand, which lie lower in the energy
spectrum, represent excitations followed by reversal of electron spins
in the system. The vortex current-density is given by \ref{vorcur} where
again $\alpha$ is the phase of the bosonic field of the standard
Chern-Simons formulation, and the skyrmion current-density \cite{leeka,stone}
 is:
\begin{equation}
{\cal J}_{\mu}^{s}
 = \frac{1}{2 \pi i} \epsilon_{\mu \nu \lambda} 
\partial^{\nu} \overline{z} \partial^{\lambda} z.
\end{equation}
The field $z$ is the two component spinor of the standard Chern-Simons
formulation \cite{leeka}, in which the bosonic field is decomposed in an amplitude,
phase, and spinor part:
\begin{equation}
\Psi_{\rm bosonic} = \rho \; \exp\{i \alpha\} \; z \label{field}
\end{equation}
The spinor part describes the spin degree of freedom of the bosonic field 
associated with the original 
 electron field.
  
The fourth, extra term with respect to spinless case in the Lagrangian
density is the nonlinear sigma model term with $ \vec{n} = \overline{z}
\vec{\tau} z $, where $ \tau_{x}, \tau_{y},$ and $\tau_{z}$ are Pauli
matrices. $ \rho_{s}$ is the stiffness constant. This term represents
the cost in the exchange energy when the ground state ferromagnetic
configuration is modified.

The fifth term is the Zeeman term with constant $ \Delta = \frac{g
\mu_{B}}{2 \pi} B $ where $B$ is the external magnetic field.

To get the low-energy, effective theory which includes the edge physics,
we repeat steps that we described in the previous subsection in the
spinless case. The constraint equation in this case, as we vary $ A_{0}$
is:
\begin{equation}
\frac{1}{2 \pi} (m \vec{\nabla} \times \vec{A} - \vec{\nabla} \times
\vec{A}^{ext}) = - {\cal J}_{0}^{v} - {\cal J}_{0}^{s}.
\end{equation}
The solution to this constraint is:
\begin{equation}
\delta A_{a} = A_{a} - \frac{1}{m} A_{a}^{ext} = - \frac{1}{m}
\partial_{a} \alpha + \frac{i}{m} \overline{z} \partial_{a} z
\label{ssolu}
\end{equation}
Now we assume that $\alpha$ is analytic so that no vortex excitations in
the bulk are allowed. We do not make any restrictions on $z$.
By plugging in the solution \ref{ssolu} in the begining Lagrangian
\ref{lagrs} with $ A_{0}=0 $ and without the second and the last term we
get surface terms:
\begin{equation}
{\cal L}^{\rm kin}_{\rm eff}(x,t)
 = \frac{1}{4 \pi m}(\partial_{x} \alpha 
 \partial_{0} \alpha -2 i \partial_{x} \alpha (\overline{z} \partial_{0}
z) + {\rm terms \; without} \; \alpha) \label{surter}
\end{equation}
The field $\alpha$ can not be found anymore in the bulk Lagrangian.
Therefore pure charge degrees of freedom i.e. those that are not
followed by a change in the spin configuration are now restricted to
live only on the boundary of the system. (This coincides with the
microscopic physical picture that we have of these systems. The pure
charge (quasihole) excitations, which lie higher in the energy spectrum
than skyrmion excitations, in the bulk, can be found in the low enery
approximation on the edge of the system. There, on the edge, their
excitation energy is smallest (in fact goes to zero).\cite{mini}) In \ref{surter} we see the most
important result of this derivation, namely a nontrivial coupling
between the spin and charge degrees of freedom on the edge of the
system.

In the low-energy approximation that we make (i.e. neglecting the last term in
\ref{lagrs}) we find that the charge current is equal to the sum of
skyrmion and vortex currents. As we do not have vortex currents in the
bulk of the system, the condition that the skyrmion current normal to
the boundary is zero is equvalent to the demand that the charge current
normal to the boundary is zero. In the absence of any external
electromagnetic fields besides the uniform magnetic that defines the
problem, we will require that no charge can leave the system i.e. the
skyrmion current normal to the boundary is zero. In that case
\begin{equation}
\partial_{x}(\overline{z} \partial_{0} z) = \partial_{0} (\overline{z}
\partial_{x} z)
\end{equation}
and we may rewrite (after a partial integration) the second surface term
in the ${\cal L}_{\rm eff}^{\rm kin}(x,t)$ as
\begin{equation}
\frac{1}{4 \pi m} [-i \partial_{x} \alpha (\overline{z} \partial_{0} z)
- i \partial_{0} \alpha (\overline{z} \partial_{x} z)]
\end{equation}
Besides
, as we did not break the gauge invariance under the gauge
transformation defined as:
\begin{equation}
\alpha \rightarrow \alpha + \beta \;\;{\rm and} \; \;z \rightarrow \exp\{-i
\beta\} \; z
\end{equation}
and present in the bosonic Chern-Simons formulation with field \ref{field},
we may also demand the same invariance on the edge of the system. This
invariance is an expression of the confinement of spin and charge on
electrons, and should exist also on the edge. As a
result the surface gauge-invariant kinetic term that contains the field
$\alpha$ is
\begin{equation}
{\cal L}_{\rm eff}^{\rm kin}=
 = \frac{1}{4 \pi m} (\partial_{x} \alpha - i 
\overline{z} \partial_{x} z) (\partial_{0} \alpha - i \overline{z}
\partial_{0} z).
\end{equation}
We can conclude that with respect to the spinless case instead of
\begin{equation}
\partial_{\mu} \alpha \; \; \mu = 0,x, \label{exp}
\end{equation}
in the case with spin we have
\begin{equation}
\partial_{\mu} \alpha - i \overline{z} \partial_{\mu} z \; \; \mu = 0,x.
\label{expres}
\end{equation}
By considering couplings with external fields we may also conclude that
these
gauge-invariant expressions  up to some appropriate
constants represent charge density and current on the edge 
similarly  to the
 the spinless case. Then \ref{expres} formally expresses the physical
fact that the charge current and density on the edge have contributions
followed by changes in the spin configuration on the edge. Therefore we expect that
under inclusion of a charge-density interaction term (which describes
the dynamics on the edge) so that the Lagrangian density is
\begin{equation}
{\cal L}_{\rm eff}
 = \frac{1}{4 \pi m} (\partial_{0} \alpha - i \overline{z}
\partial_{0} z) (\partial_{x} \alpha - i \overline{z} \partial_{x} z)
- \frac{v}{4 \pi m} (\partial_{x} \alpha - i \overline{z} \partial_{x}
z)^{2}, \label{rezultat}
\end{equation}
we have a complete low-energy description of the charge degree of
freedom on the edge. The equation of motion for $\alpha$ on the edge
simply tells us that the charge on the edge drifts with velocity $v$
along the edge in only one direction as it should be in a quantum Hall
system.
    
At this stage of deriving the low-energy, effective theory, the bulk
Lagrangian density is
\begin{equation}
\Delta {\cal L}_{\rm eff}(x,y,t)
 = - {\cal J}_{a}^{s} (\frac{1}{m} A_{ext}^{a}+\frac{i}{m}
\overline{z} \partial^{a} z) -
\frac{\rho_{s}}{2} (\vec{\nabla} \vec{n})^{2} + \frac{\Delta}{2} n_{3}
\label{restlag}
\end{equation}
where $ a = x,y $. Now we will 
 apply the spin-wave approximation in which field $z$ is
decomposed in the following way
\begin{equation}
z= \left(\begin{array}{c}
       1- \frac{1}{2} |\Psi|^{2} \\ \Psi
         \end{array} \right).
\end{equation}
The complex bosonic field $\Psi$ expresses fluctuations with respect to
the ground state configuration and is to be considered small and slowly
varying, and therefore, we will keep only terms quadratic in $\Psi$.
Then, the $2+1$ dimensional Lagrangian density \ref{restlag}
of the bulk becomes
\begin{equation}
{\cal L}_{\rm eff}(x,y,t)
 = i \rho_{0} \overline{\Psi} \partial_{0} \Psi
- 2 \rho_{s} \vec{\nabla} \overline{\Psi} \vec{\nabla} \Psi - \Delta
|\Psi|^{2} \label{dvapet}
\end{equation}
Now, we will again assume that our system is in the lower half-plane.
After a partial integration inside the expression for the system
Lagrangian we get an extra (to the chiral boson Lagrangian) surface
term:
\begin{equation}
\Delta {\cal L}_{\rm surf}(x,t)
 = - 2 \rho_{s} \overline{\Psi} \partial_{y}
\Psi,   \label{extrasur}
\end{equation}
and the $2+1$ dimensional Lagrangian density of the bulk
\begin{equation}
{\cal L}_{\rm eff}^{\rm bulk}(x,y,t)
 = i \rho_{0} \overline{\Psi} \partial_{0}
\Psi + 2 \rho_{s} \overline{\Psi} \vec{\nabla}^{2} \Psi - \Delta
|\Psi|^{2}. \label{lagbu}
\end{equation}
For a moment we will discuss only the problem of 2D ferromagnet
described by the Lagrangian densities \ref{extrasur} and \ref{lagbu} (in
the spin-wave approximation) in order to explain the meaning of the
surface term \ref{extrasur}.
In that case, the corresponding Euler-Lagrange equations for the field
$\Psi$ is
\begin{equation}
i \rho_{0} \partial_{0} \Psi + 2 \rho_{s} \vec{\nabla}^{2} \Psi - \Delta
\Psi = 0 \label{bueq}
\end{equation}
in the bulk, and
\begin{equation}
\partial_{y} \Psi|_{y=0} = 0 \label{baeq}
\end{equation}
on the edge. The normal mode solutions of \ref{bueq} in the bulk are
spin waves,
\begin{equation}
\Psi = A \exp\{ i \vec{k} \vec{r}\} \exp\{ - i w t\}
\end{equation}
with the dispersion relation:
\begin{equation}
w = \frac{\Delta}{\rho_{0}} + \frac{2 \rho_{s}}{\rho_{0}} k^{2}.
\end{equation}
To satisfy boundary condition \ref{baeq}, the class of solutions is
further reduced to the form:
\begin{equation}
\Psi = A \cos\{ k_{y} y\} \exp\{i k_{x} x\} \exp\{- i w t\}.
\label{spwe}
\end{equation}
The condition \ref{baeq} ensures that spin currents normal to the
boundary are zero. Namely, if we use the Noether expression for them,
\begin{equation}
{\cal J}_{\mu}^{a}
 = - \frac{i}{2} 
\frac{\delta
{\cal L}_{QFM}}
{\delta(\partial^{\mu} z_{i})} \tau_{ij}^{a} z_{j} + {\rm
h.c.}
\end{equation}
where $ \tau^{a}, a = 1,2,3 $ are Pauli matrices and ${\cal L}_{QFM}$ (a
Lagrangian density of a quantum ferromagnet) is
\begin{equation}
{\cal L}_{QFM}
 = i \rho_{0} \overline{z} \partial_{0} z -
\frac{\rho_{s}}{2} (\vec{\nabla} \vec{n})^{2} + \frac{\Delta}{2}
\overline{z} \tau_{3} z,
\end{equation}
we may find in the spin-wave approximation the following expressions for
them:
\begin{eqnarray}
& & {\cal J}_{y}^{3} = i \rho_{s} (\overline{\Psi} \partial_{y} \Psi -
\partial_{y} \overline{\Psi} \Psi)  \nonumber \\
& & {\cal J}_{y}^{1} = i \rho_{s} (\partial_{y} \overline{\Psi} -
\partial_{y} \Psi \nonumber \\
& & {\cal J}_{y}^{2} = i \rho_{s} (\partial_{y} \overline{\Psi} +
\partial_{y} \Psi). \label{spcurrents}
\end{eqnarray}
We neglected the terms of order higher than two in $\Psi$; the single
condition \ref{baeq} ensures that all of them are zero at the boundary.

The edge spin-wave solutions of the form
\begin{equation}
\Psi = B \exp\{\beta y\} \exp\{i k_{x} x\} \exp\{-i w t\}, \label{edwave}
\end{equation}
$\beta$ positive, satisfy the equation in the bulk, but the condition
\ref{baeq} forces $\beta$ to be zero. Therefore, in the case of a pure
quantum ferromagnet, the constraint of the spin conservation on the
boundary, does not allow the existance of the edge spin excitations.

Now, we will summarize our effective theory for a quantum Hall system with
a boundary, having in mind the specific geometry in which the system is
in the lower half of the plane. The bulk Lagrangian density (in the
spin wave approximation) is given by
\begin{equation}
{\cal L}_{\rm bulk}
 = i \rho_{0} \overline{\Psi} \partial_{0} \Psi +
2 \rho_{s} \overline{\Psi} \vec{\nabla}^{2} \Psi - \Delta |\Psi|^{2}+
{\rm higher \; order \; terms} \label{effbu}
\end{equation}
and the edge Lagrangian density is
\begin{equation}
{\cal L}_{\rm edge}(x,t)
 = - 2 \rho_{s} \overline{\Psi} \partial_{y} \Psi +
\frac{1}{4 \pi m}( \partial_{x} \alpha^{f} \partial_{0} \alpha^{f} - v
 \partial_{x} \alpha^{f} \partial_{x} \alpha^{f}) + {\rm higher \; order
\;
terms}   \label{effed}
\end{equation}
where
\begin{equation}
\partial_{\mu} \alpha^{f} = \partial_{\mu} \alpha - \frac{i}{2}
(\overline{\Psi} \partial_{\mu} \Psi - \partial_{\mu} \overline{\Psi}
\Psi).
\end{equation}

\subsection{The proof of gauge invariance of the effective field theory}

In the spinless case the effective action of the bulk,
\begin{equation}
{\cal L}_{\rm bulk} = - \frac{m}{4 \pi} \epsilon^{\mu \nu \lambda}
A_{\mu} \partial_{\nu} A_{\lambda} + \frac{1}{2 \pi} \epsilon^{\mu \nu
\lambda} A^{ext}_{\mu} \partial_{\nu} A_{\lambda} \label{pocetak}
\end{equation}
when the field $ A_{\mu}$ is integrated out, is
\begin{equation}
{\cal L}_{\rm bulk}^{\rm eff} = \frac{1}{4 \pi m} \epsilon^{\mu \nu \lambda}
A_{\mu}^{ext} \partial_{\nu} A^{ext}_{\lambda}.
\end{equation}
On the other hand the edge action with external electromagnetic field
included is \cite{wen,ston}
\begin{eqnarray}
{\cal L}_{\rm edge}= & & \frac{1}{4 \pi m} [\partial_{x} \alpha
\partial_{0} \alpha - v (\partial_{x} \alpha)^{2}]  \nonumber \\
& & \frac{1}{2 \pi m} (v A_{x}^{ext} - A_{0}^{ext}) \partial_{x} \alpha 
- \frac{1}{4 \pi m} (v A_{x}^{ext} - A_{0}^{ext}) A_{x}^{ext}
\label{edaction}
\end{eqnarray}
Under the gauge transformation $ \alpha \rightarrow \alpha + \Lambda $
and $ A_{\mu}^{ext} \rightarrow A_{\mu}^{ext} + \partial_{\mu} \Lambda$
the total action, $ {\cal L}_{\rm bulk} + {\cal L}_{\rm edge}$ or
$ {\cal L}_{\rm bulk}^{\rm eff} + {\cal L}_{\rm edge}$, is
invariant. Namely the chiral anomaly \cite{ston} term
\begin{equation}
\frac{1}{4 \pi m} \Lambda (\partial_{0} A_{x}^{ext} - \partial_{x} A_{0}^{ext})
\end{equation}
which we get by the gauge transformation of the bulk action is of the
same absolute value but of opposite sign as the term that we get gauge
transforming the edge action \ref{edaction}.

In our previous derivation we assumed that only the vector potential of
the constant magnetic field is present in \ref{pocetak}
 and, therefore, the field $ A^{\mu}$ did not have any dynamics; by the
equation of motion of the action it was constrained to satisfy $ m
\epsilon^{a b} \partial_{a} A_{b} = \epsilon^{a b} \partial_{a}
A_{b}^{ext} $. Because of the absence of the perturbing external
electromagnetic fields the whole dynamics of the system was on the edge
described by the chiral boson theory (the first two terms in
\ref{edaction}).

Similarly, in the case with spin, more general effective bulk action
with external perturbing electromagnetic fields is
\begin{eqnarray}
{\cal L}_{\rm bulk} = & & - \frac{m}{4 \pi} \epsilon^{\mu \nu \lambda}
 A_{\mu} \partial_{\nu} A_{\lambda} + \frac{1}{2 \pi} \epsilon^{\mu \nu
\lambda} A_{\mu}^{ext} \partial_{\nu} A_{\lambda} - {\cal J}^{s}_{\mu}
A_{\mu}  \nonumber \\
& & - \frac{\rho_{s}}{2} (\vec{\nabla} \vec{n})^{2} + \frac{\Delta}{2}
n_{3}. \label{begin}
\end{eqnarray}
Compare with \ref{lagrs} and note the absence of the vortex current
${\cal J}^{v}_{\mu}$ in the action. As we pointed out before this
signifies the fact that the charge excitations without spin changes are
to be found on the edge of the system in the low-energy approximation.

If we integrate out $A_{\mu}$ field we get
\begin{equation}
{\cal L}_{\rm bulk}^{\rm eff} = \frac{1}{4 \pi m} \epsilon^{\mu \nu \lambda}
A_{\mu}^{ext} \partial_{\nu} A_{\lambda}^{ext} + \frac{1}{4 \pi m}
\epsilon^{\mu \nu \lambda} A_{\mu}^{ext} \partial_{\nu}(i \overline{z}
\partial_{\lambda} z) + {\rm terms \; \; without \; \;} A_{\mu}^{ext}
\label{spinpocetak}
\end{equation}
For the edge the complete action (see \ref{rezultat}) with external
electromagnetic fields is
\begin{eqnarray}
{\cal L}_{\rm edge} = & & \frac{1}{4 \pi m} [(\partial_{x} \alpha - i
\overline{z} \partial_{x} z) (\partial_{0} \alpha - i \overline{z}
\partial_{0} z) - v (\partial_{x} \alpha - i \overline{z} \partial_{x}
z)^{2}] \nonumber \\
& &
+ \frac{1}{2 \pi m} (v A_{x}^{ext} - A_{0}^{ext}) (\partial_{x} \alpha - i
\overline{z} \partial_{x} z) - \frac{1}{4 \pi m} (v A_{x}^{ext} - A_{0}^{ext}
)A_{x}^{ext} \label{xxaction}
\end{eqnarray}
Again, in this case, the extra term that we get by setting
$A_{\mu}^{ext} \rightarrow A_{\mu}^{ext} + \partial_{\mu} \Lambda $ in
\ref{spinpocetak},
\begin{equation}
\frac{1}{4 \pi m} \Lambda [\partial_{0}(A_{x}^{ext} + i \overline{z}
\partial_{x} z) - \partial_{x}(A_{0}^{ext} + i \overline{z} \partial_{0}
z)]
\end{equation}
is canceled by the term that comes out from taking $ A_{\mu}^{ext}
\rightarrow A_{\mu}^{ext} + \partial_{\mu} \Lambda$ and $\alpha
\rightarrow \alpha + \Lambda$ in \ref{xxaction}. Therefore the gauge
invariance of the total action, $ {\cal L}_{\rm bulk} + {\cal L}_{\rm
edge}$ (or ${\cal L}^{\rm eff}_{\rm bulk} + {\cal L}_{\rm edge}$, \ref{begin} and \ref{xxaction}, is proved.

As we take $ A_{\mu}^{ext}$ to be the vector potential of the constant
magnetic field and apply the spin-wave approximation the action
\ref{begin} is transformed into \ref{dvapet} and the edge action becomes
\ref{rezultat}.

\section{Solutions of the field theory}
As we vary the surface terms of the low-energy effective Lagrangian 
(
\ref{effbu} and \ref{effed})with respect to $\alpha$ and
$\overline{\Psi}$ (or $\Psi$) we get two equations.
As we vary $\alpha$, we get
\begin{equation}
\partial_{x} \partial_{0} \alpha^{f} = v_{c} \partial_{x}^{2} \alpha^{f}
\label{chargeeq}
\end{equation}
and, by varying $\overline{\Psi}$ and using the  previous equation,
\begin{equation}
2 \rho_{s} \partial_{y} \Psi + \frac{1}{4 \pi m} (-i) [ \partial_{x}
\Psi \partial_{0} \alpha^{f} - \partial_{0} \Psi \partial_{x} \alpha^{f}
- v_{c} 2 \partial_{x} \Psi \partial_{x} \Psi \partial_{x} \alpha^{f}]=0
\label{boundarye}
\end{equation}
When $\partial_{\mu} \alpha^{f} = 0, \mu =0,x$ i.e. there are no charge
excitations on the boundary, the spin waves \ref{spwe} are solutions of
the bulk and surface equations. When $\Psi = 0$, i.e. there are no spin
excitations in the system; the only solutions of the equations are
charge density waves of the chiral boson theory.

From the bulk equation \ref{bueq} we get the following dispersion
relation for the edge spin waves \ref{edwave}:
\begin{equation}
w = \frac{\Delta}{\rho_{0}} + \frac{2 \rho_{s}}{\rho_{0}} (k_{x}^{2} -
\beta^{2})    \label{disprel}
\end{equation}
The coefficient $\beta$, which characterize the extension of the edge
spin waves into the bulk of the system, comes from the boundary equation
\ref{boundarye}. The one-dimensional charge density and current are
given by
\begin{equation}
j_{0} = \frac{1}{2 \pi m} \partial_{x} \alpha^{f} \;\; {\rm and} \;\; 
j_{x} = \frac{1}{2 \pi m} \partial_{0} \alpha^{f}
\end{equation}
respectively. For $\alpha^{f} = \alpha^{f}(x + v_{c}t)$, the general
solution of \ref{chargeeq}, we have
\begin{equation}
v_{c} \partial_{x} \alpha^{f} = \partial_{0} \alpha^{f} \; \; {\rm i.e.}
\; \;  v_{c} j_{0} = j_{x}
\end{equation}
and we may rewrite the equation \ref{boundarye} as
\begin{equation}
- 4 \rho_{s} \partial_{x} \Psi + i j_{0} \partial_{0} \Psi + i j_{0}
v_{c} \partial_{x} \Psi = 0 \label{spbaeq}
\end{equation}

\subsection{Charged edge spin-wave solutions}
The solution of the form \ref{edwave} of the equation \ref{spbaeq} with $\beta =$
const exists only if $j_{0}$ is, by itself, a constant, i.e. if there is
a constant charge density along the boundary of the system. For the
ground state $j_{0} = 0$, and the previous condition (with $j_{0} \neq
0$) means that some (extra) charge is added to $(j_{0}>0)$ or subtracted
from the system $(j_{0}<0)$. Plugging in the form \ref{edwave} into
\ref{spbaeq}, and using \ref{disprel}, we obtain the following
candidates for solutions with $\beta$ equal to
\begin{equation}
\beta_{1,2} = - \frac{\rho_{0}}{j_{0}} \pm \frac{\rho_{0}}{|j_{0}|}
\sqrt{1 + \frac{j_{0}^{2}}{2 \rho_{s} \rho_{0}^{2}} (\Delta +
2 \rho_{s} k_{x}^{2} - v_{c} \rho_{0} k_{x})} \label{betasol}
\end{equation}
from which only those with $\beta$ positive can describe the edge
spin-waves. In the approximations in which  the
second term under the square root in \ref{betasol} is small,
\begin{eqnarray}
\beta \approx & \frac{j_{0}}{4 \rho_{s} \rho_{0}} (\Delta + 2 \rho_{s}
k_{x}^{2} - v_{c} \rho_{0} k_{x}) & \; \;{\rm for}\; \; j_{0}> 0   \nonumber \\
{\rm and}\; \beta \approx & 2 \frac{\rho_{0}}{|j_{0}|} & \; \;{\rm for}
\; \;j_{0} <
0 \label{twoeq}
\end{eqnarray}
For $j_{0} \rightarrow 0$ the second solution is unacceptable because
our effective theory assumes (spatially) slowly varying quantities,
which is not the case with the solution with $\beta$ large.
 
Also, we may notice looking at \ref{twoeq} asymmetry between $j_{0}>0$
and $j_{0}<0$ case. One way to understand the asymmetry is to first
consider the single particle picture of adding and subtracting charge to
the edge of these systems. In this picture of systems without spin,
adding or subtracting charge is always equivalent to simple shifts
(translations) of the boundary. In the case of the systems with spin
there is an additional possibility to add charge as shown in Fig.1.

We believe that this single particle picture is behind the many body
state - charged edge spin-wave for $j_{0}>0$. This can be supported by
the following consideration. First, we may try to interpret the solution
as a solution of a generalized pure-spin (quantum-ferromagnet) problem,
in which the second and third term in \ref{spbaeq} correspond to some
surface terms in that problem. For a finite Zeeman coupling and in the
small-momentum approximation, the solution with $j_{0}>0$ corresponds to
the following term:
\begin{equation}
\frac{j_{0}}{\rho_{0}} \Delta \overline{\Psi} \Psi
\end{equation}
with an effective magnetic field $ B_{eff} = -(\frac{j_{0}}{\rho_{0}}) B $
on the boundary. As we add electrons, we effectively create a magnetic
field in the opposite direction of the external magnetic field. This
magnetic field makes possible the creation of the spin edge solution. As
we add more electrons ($j_{0}$ larger) the effective magnetic field
increases in its magnitude and favors edge spin waves localized near the
boundary ($\beta$ larger). Therefore the edge-spin-wave, one-spin flip
will be more localized on the boundary  if there are more pairs
(see Fig.1) each of them energetically unfavorable because each of them
consists of two electrons in the same orbital. 

The corresponding term (in the pure-spin problem) for the second
solution (with $j_{0} < 0$) is
\begin{equation}
- \frac{2 \rho_{s} j_{0}}{\rho_{0}} \partial_{y} \overline{\Psi}
\partial_{y} \Psi.
\end{equation}
It is of the opposite sign than the gradient term in the bulk and,
therefore, favors the solution which disorders the spin configuration of
the ground state and, we expect as it is usual in these systems, that it
is also followed by a redistribution of the charge on the boundary.

We also expect that these charged edge spin-waves can be created without
any change in the total charge of the system i.e. by simple
redistributions of the ground-state  charge on the boundary where
$j_{0}$ is  a parameter that characterizes them. Therefore, when
considering the total energy of excitations we will neglect the energy
term of the chiral boson Lagrangian. Then the surface
contribution is
\begin{equation}
E_{\rm surface} = 2 \rho_{s} \beta^{2}  \label{ensur}
\end{equation}
and the contribution from the bulk is given by \ref{disprel}.(To get
\ref{ensur} we normalized the wave to describe one electron spin flip:
\begin{equation}
\Psi = \sqrt{\frac{\beta}{\pi}} \exp\{ \beta y + i k x - i w t\})
\end{equation}
 As a
consequence, we may write the total energy of excitations as
\begin{equation}
E_{tot} = \frac{\Delta}{\rho_{0}} + \frac{2 \rho_{s}}{\rho_{0}} k^{2} +
\beta^{2} (4 \rho_{s} - \frac{2 \rho_{s}}{\rho_{0}}) \label{etot}
\end{equation}
Because $ \rho_{0} = \frac{1}{2 \pi m}$ the last term in \ref{etot} is
always negative, and at $k=0$ $E_{tot}$ is always lower than the
constant Zeeman term. When \ref{twoeq} is substituted for $\beta$ in the
expression \ref{etot}, the energy \ref{etot} is of the following form
\begin{equation}
E_{tot} = c_{0} + c_{1} k + c_{2} k^{2}+ \cdots
\end{equation}
(where $c$'s do not depend on k). By having a linear term in $k$ this
dispersion relation is asymmetric around $k=0$. 

\subsection{Neutral edge spin-wave solutions}
We define the neutral edge spin-waves by requiring that the field
$\alpha$, which lives on the boundary, associated with pure (not carryng
also spin) charge degrees of freedom, is zero ($\alpha = 0$). As a
consequence there is one less surface equation to be satisfied, because
 the constraint equations \ref{chargeeq} and \ref{boundarye} can not
be satisfied simultaneously. And, as we will see, the constraint also
implies losing  one of the two (charge and spin) local conservation
laws. Namely, local spin currents normal to the boundary will be
nonzero in general. It will be also shown that these excitations are
neutral i.e. the total change in the charge of the system, when these
excitations occur, is zero.

With the condition $\alpha = 0$ the charge current and density on the
boundary are proportional to
\begin{equation}
\partial_{\mu} \alpha^{f} = - \frac{i}{2} (\overline{\Psi}
\partial_{\mu} \Psi - \partial_{\mu} \overline{\Psi} \Psi) \;\;
\mu = 0,x
\end{equation}
These charge degrees of freedom must satisfy \ref{chargeeq}, and from
that it follows that the dispersion relation for the waves is
$ w = v_{c} k$. This dispersion relation follows when a general solution
that is a superposition of the waves of the form \ref{edwave} is
considered. The frequency of the wave is also equal to \ref{disprel}.
This enables us to find $\beta$ in this case. It is given by the
following expression
\begin{equation}
\beta^{2} = \frac{\Delta}{2 \rho_{s}} + k^{2} - \frac{v_{c} \rho_{0}}{2
\rho_{s}} k,  \label{betav}
\end{equation}
i.e. in the small-momentum approximation $ \beta \approx
\sqrt{\frac{\Delta}{2 \rho_{s}}}$.
But the waves do not satisfy \ref{boundarye}, which, in the spin-wave
approximation, is equivalent to the condition \ref{baeq}. This means
that the spin currents normal to the boundary of the system are, in
general, nonzero and exchange of the spin of the system with outside is
allowed. Nevertheless, ${\cal J}_{y}^{3} = 0$ (see \ref{spcurrents})
everywhere on the boundary and, also, the total spin of the system is
conserved, i.e.
\begin{equation}
 \int_{-\infty}^{+\infty}  {\cal J}_{y}^{2} dx = 0 \;\; {\rm and}
\;\; \int_{-\infty}^{+\infty} {\cal J}_{y}^{1} dx = 0
\end{equation}

Although the bulk excitation energy $(w = v_{c} k)$ is gapless, these
excitations have a gap. To see this we have to calculate their total
energy and include the surface contribution \ref{ensur}, with $\beta$
given by \ref{betav}. As a final result we have:
\begin{equation}
E_{tot} = 2 \Delta + v_{c} k (1 - 2 \rho_{s}) + 4 \rho_{s} k^{2}
\label{nuttot}
\end{equation}
The requirement $E_{tot}>0$ for each k fixes the allowed range of the
parameters in \ref{nuttot} and in the theory. The excitations are
allowed to propagate in both directions, though, the spectrum is
assymetric around $k = 0$ because of the presence of the linear term in
$E_{tot}$. The gap (the smallest excitation energy for certain $k$) of
these excitaations is always smaller than the Zeeman gap.

To prove the neutrality of these excitations we start with a general
edge spin-wave solution:
\begin{equation}
\Psi(x,y,t) = \sum_{k} \sqrt{\frac{\beta}{\pi}} 
\exp\{ \beta y + i k x - i w t\} a(k),
\end{equation}
where $a(k)$ are arbitrary (complex) coefficients in this expansion. The
density of charge on the boundary of the system can be expressed as
\begin{eqnarray}
\rho_{\rm surface}(x,t) & = & \frac{\partial_{x} \alpha^{f}}{2 \pi m} =
\frac{1}{2 \pi m} (-\frac{i}{2}) (\overline{\Psi} \partial_{x} \Psi -
\partial_{x} \overline{\Psi} \Psi) = \nonumber \\
& = & \frac{\beta}{2 \pi^{2} m} (-\frac{i}{2}) \sum_{k,k^{\prime}}
\exp\{-i (k - k^{\prime})x\}
\exp\{i(w_{k} - w_{k^{\prime}})t\}
i (k^{\prime} + k) a(k)^{\ast} a(k^{\prime})
\end{eqnarray}
The same quantity in the bulk is proportional to the topological density
 \cite{leeka} and equal to
\begin{equation}
\rho_{\rm bulk}(x,y,t) = \frac{(-i)}{2 \pi m} (\partial_{x} \overline{z}
\partial_{y} z - \partial_{y} \overline{z} \partial_{x} z).
\end{equation}
In the spin-wave approximation we have
\begin{eqnarray}
\rho_{\rm bulk}(x,y,t) &  = & \frac{(-i)}{2 \pi m} (\partial_{x} \overline{\Psi}
\partial_{y} \Psi - \partial_{y} \overline{\Psi} \partial_{x} \Psi)
\nonumber \\
  &  =  & \frac{-i \beta^{2}}{2 \pi^{2} m} \sum_{k,k^{\prime}}
\exp\{2 \beta y\}
\exp\{-i (k - k^{\prime})x\}
\exp\{i (w_{k} - w_{k^{\prime}})t\}
(-i) (k + k^{\prime}) a(k)^{\ast} a(k^{\prime}).
\end{eqnarray}
Clearly, the total change in the charge of the boundary,
\begin{equation}
Q_{\rm surface} = \int_{-\infty}^{+\infty} \rho_{\rm surface}(x,t) dx =
\frac{\beta}{\pi m} \sum_{k} k a(k)^{\ast} a(k),
\end{equation}
is of the same amount, but of opposite sign than the one in the bulk
$(Q_{\rm bulk} = \int_{-\infty}^{+\infty} dx \int_{-\infty}^{0} dy
 \rho_{\rm bulk}(x,y,t))$.

At the end of this section we would like to comment on the nature of
these neutral edge spin waves. As they are completely skyrmionic, their
charge is fully specified by their spin configuration. Their spread
increases with the decreasing of the Zeeman energy, which is well known
skyrmion property. As they carry fixed (unit) amount of spin they vanish
when the Zeeman coupling is zero. Due to the close relationship between
their charge and spin and linearity of the chiral boson dispersion
relation, the dispersion relation of these spin waves is also linear.

\section{Conclusions and Discussion}
In conclusion, we proposed an effective edge theory for ferromagnetic quantum
Hall states. It describes their (1 + 1) dimensional charge degrees of freedom
by a chiral boson theory  and their (2 + 1) dimensional spin
degrees of freedom by the effective theory of a quantum ferromagnet 
 in the spin-wave approximation.
 We found two classes of the
edge spin-wave solutions. The class with the charged edge spin waves is obtained by
removing or adding electrons to the edge of the system. 
 The second class of the neutral edge spin waves does not
require any change in the total charge of the system. All these edge
spin excitations are characterized by
 linear dispersion relations and gaps in the excitation energy.

We did not consider most-general surface terms for the spin degrees of freedom
which would be present in a most-general theory, because we wanted to
emphasize and examine the influence of the terms that describe  the charge degrees of
freedom. The latter terms, collected in the chiral boson theory, give
a complete effective description of the charge degrees of freedom on the edge.

Ref. \cite{kkls} pointed out that for small Zeeman energies and soft confining
potentials a reconstruction (from narrow and spin polarized edge) to the
edge with spin textures should occur.
% Despite our assumption of the
%completely polarized ground state with a sharp edge, we expect that our effective
%theory is valid even in that case. Namely, in our theory the width of the
%edge relevant to the charge degrees of freedom is zero, because the 
%characteristic width
%for the edge spin degrees of freedom that we wanted to describe is expected to be
%much larger than the one for the charge. In that case the chiral boson theory
%captures the essential physics for the charge degrees of freedom on the edge,
%even in the presence of the edge reconstruction whose characteristic width
%(see for estimates in \cite{kkls}) becomes zero.
Because of our assumption of the edge with the same polarization as
that of
the bulk, our theory is valid only for steep enough (see for
estimates in \cite{kkls}) confining potentials.

At the end we would like to comment on the relationship between our work
and the recent work concerning edge excitations and edge reconstruction
at $\nu=$ 1 \cite{kkll}. There, the Hartree-Fock procedure was used to
determine the energy spectrum of some proposed edge spin-flip
excitations. (In fact a special form of these excitations was first
suggested in Ref. \cite{ommt}.) These excitations reconstruct the edge
when the Zeeman energy is small and the confining potential softened.
They are followed by an outward movement of charge and exist even at
zero Zeeman energy. The latter property is not shared by our neutral
edge excitations and charged edge excitations for $j_{0}>0$ and
therefore, they should not be confused with the ones of Ref. \cite{kkll}.
(The neutral edge waves even require, as we described, special boundary
conditions to exist.) On the other hand the detailed description of the
charged edge excitations for $ j_{0}<0 $ (which can be associated with
the outward movement of charge) is impossible in the scope of our
theory. The theory gives only an inkling of the possible instability of
the ground state with respect to their creation.

The author thanks F.D.M. Haldane and N. Read for beneficial discussions.
She is especially thankful to E. Shimshoni who also gave very useful
comments on the manuscript. 

This work was supported by Israel Council
for Higher Education.

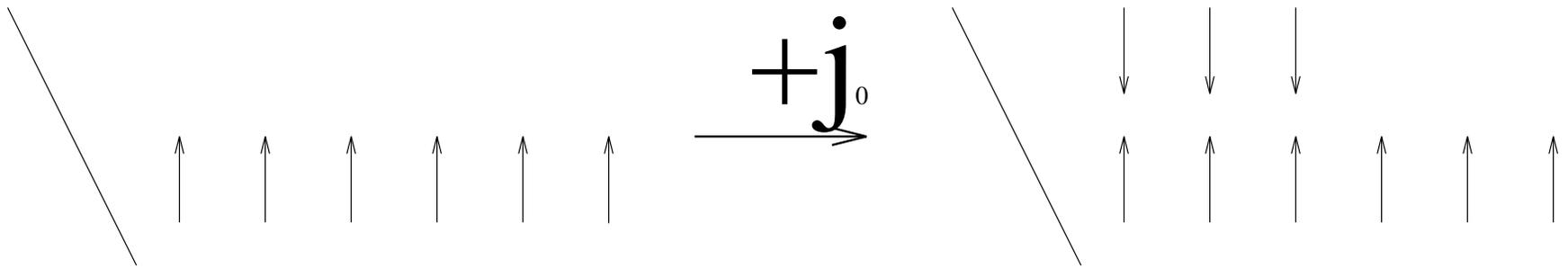
\begin{figure}
\caption{Adding  charge to the edge of $ \nu = 1$ filling fraction in
single-particle picture}
\end{figure}

\begin{references}
\bibitem{wen}
X.-G.\ Wen, Phys. Rev. B {\bf 41}, 12838 (1990); {\bf 43}, 11025 (1991);
Int. J. Mod. Phys. {\bf B6}, 1711 (1992).
\bibitem{sond}
S.L.\ Sondhi, A.\ Karlhede, S.A.\ Kivelson, and E.H.\ Rezayi, 
Phys. Rev. B {\bf 47}, 16419 (1993).
\bibitem{barr}
S.E.\ Barrett, G.\ Dabbagh, L.N.\ Pfeifer, K.W.\ West, and R.\ Tycko,
Phys. Rev. Lett. {\bf 74}, 5122 (1995).
\bibitem{ston}
M.\ Stone, Ann. of Phys. {\bf 207}, 38 (1991).
\bibitem{hald}
F.D.M.\ Haldane, Phys. Rev. Lett. {\bf 74}, 2090 (1995).
\bibitem{leeka}
D.H.\ Lee and C.L.\ Kane, Phys. Rev. Lett. {\bf 64}, 1313 (1990).
\bibitem{stone}
M.\ Stone,  cond-mat/9512010.
\bibitem{mini}
M.\ Milovanovi\'{c} and N.\ Read, Phys. Rev. B {\bf 53}, 13559 (1996).
\bibitem{kkls}
A.\ Karlhede, S.A.\ Kivelson, K.\ Lejnell, and S.L.\ Sondhi, 
Phys. Rev. Lett. {\bf 77}, 2061 (1996).
\bibitem{kkll}
A.\ Karlhede and K.\ Lejnell, cond-mat/9709339
\bibitem{ommt}
J.H.\ Oaknin, L.\ Martin-Moreno, and C.\ Tejedor, Phys. Rev. B {\bf 54},
16850 (1996).
\end{references}
\end{document}